**EyeDiff: text-to-image diffusion model improves rare eye disease diagnosis**

**Authors**

Ruoyu Chen[1,#], Weiyi Zhang[1,#], Bowen Liu[1], Xiaolan Chen[1], Pusheng Xu[1], Shunming Liu[1], Mingguang He[1-3*], Danli Shi[1-2*]

**Affiliation**

1. School of Optometry, The Hong Kong Polytechnic University, Kowloon, Hong Kong SAR, China

2. Research Centre for SHARP Vision, The Hong Kong Polytechnic University, Kowloon, Hong Kong SAR, China

3. Centre for Eye and Vision Research (CEVR), 17W Hong Kong Science Park, Hong Kong SAR, China

# Contributed equally

**Correspondence**

**Dr. Danli Shi**, MD, PhD., Research Assistant Professor, School of Optometry, The Hong Kong Polytechnic University, Kowloon, Hong Kong SAR, China.

Email: danli.shi@polyu.edu.hk

**Prof. Mingguang He**, MD, PhD., Chair Professor of Experimental Ophthalmology, School of Optometry, The Hong Kong Polytechnic University, Kowloon, Hong Kong SAR, China.

Email: mingguang.he@polyu.edu.hk

**Abstract**

The rising prevalence of vision-threatening retinal diseases poses a significant burden on the global healthcare systems. Deep learning (DL) offers a promising solution for automatic disease screening but demands substantial data. Collecting and labeling large volumes of ophthalmic images across various modalities encounters several real-world challenges, especially for rare diseases. Here, we introduce EyeDiff, a text-to-image model designed to generate multimodal ophthalmic images from natural language prompts and evaluate its applicability in diagnosing common and rare diseases. EyeDiff is trained on eight large-scale datasets using the advanced latent diffusion model, covering 14 ophthalmic image modalities and over 80 ocular diseases, and is adapted to ten multi-country external datasets. The generated images accurately capture essential lesional characteristics, achieving high alignment with text prompts as evaluated by objective metrics and human experts. Furthermore, integrating generated images significantly enhances the accuracy of detecting minority classes and rare eye diseases, surpassing traditional oversampling methods in addressing data imbalance. EyeDiff effectively tackles the issue of data imbalance and insufficiency typically encountered in rare diseases and addresses the challenges of collecting large-scale annotated images, offering a transformative solution to enhance the development of expert-level diseases diagnosis models in ophthalmic field.

**Introduction**

The increasing prevalence of vision-threatening retinal diseases has significantly strained the global healthcare system.[1] Multimodal ophthalmic images, such as color fundus photographs (CFP), optical coherence tomography (OCT), fundus fluorescein angiography (FFA), etc. provide complementary information on ocular abnormalities.[2-4] Deep learning (DL) techniques have shown great potential in automatically identifying lesions in these images, aiding disease diagnosis.[5-7] In this data-centric domain, the effectiveness of DL algorithms relies heavily on the availability of well-curated medical images with high-quality annotations.[8,9] However, accessing large and diverse datasets encounters several real-world obstacles, such as complex image collection procedures, concerns about patient privacy, the labor-intensive nature of human annotation, and challenges in cross-center data sharing.[10,11] Additionally, acquiring class-balanced data, particularly for rare diseases, remains a significant hurdle.

To address data scarcity and imbalance, various data augmentation techniques have been explored.[12] Traditional random oversampling rebalances classes by duplicating minority examples,[13] but this can lead to overfitting and affect model accuracy.[14] Recently, though generative adversarial networks (GANs) have been used to synthesize realistic images and improve disease diagnosis, such as diabetic retinopathy (DR) and age-related macular degeneration (AMD), most efforts have focused on unimodal images.[15-19] Besides, self-supervised learning has been explored in ophthalmology, as demonstrated by the RETFound model, to enhance data usability.[20] Nevertheless, this approach did not fully address data imbalance or the scarcity of rare disease images.

The stable diffusion (SD) model, a novel text-to-image generation approach, offers a promising solution by creating realistic images from natural language prompts.[21] In general domain, the SD can generate images that perform as well as or better than real images in training self-supervised models.[22] This approach provides a practical way to obtain diverse synthetic training data, benefiting large-scale visual representation learning. However, its application in generating multimodal ophthalmic images remains unexplored.

In this study, we innovatively introduce EyeDiff, a text-to-image diffusion model trained on extensive multimodal datasets. We validate EyeDiff by synthesizing diverse multimodal images through text prompts, supplementing specific minority classes across 10 benchmarks, and evaluating whether these generated images can enhance the diagnosis of common and rare retinal diseases

## Methods

### Datasets

EyeDiff was trained on large-scale collection of ophthalmic image-text pairs, covering 14modalities and over 80 disease categories from eight datasets. This extensive training enabled the model to accurately capture the relationships between image distributions and their corresponding textual descriptions across a broad spectrum of diseases. We then assessed the clinical value of EyeDiff by using its generated images to augment minority classes across 10 external validation datasets from various global regions, and thoroughly evaluated its performance in downstream disease diagnosis tasks. Table 1 details the characteristics of the training and external validation datasets, and the study flowchart is illustrated in Figure 1.

### Training datasets

A brief overview of the eight training datasets is as follows: (1) Retinal Image Bank: An open-access collection by the American Society of Retinal Specialists, containing over 29,000 multimodal images and descriptions covering various retinal diseases (Supplementary Table 1). (2) EyePACS: Comprising 88,702 fundus images from diverse populations with varying degrees of DR, commonly used for developing and testing DR screening models.[23] (3) OCTDL: An open-source OCT dataset with over 2,000 OCT images labeled with disease groups and retinal pathologies, including age-related macular degeneration (AMD), diabetic macular edema (DME), epiretinal membrane (ERM), retinal artery occlusion (RAO), retinal vein occlusion (RVO), and common vitreomacular interface diseases.[24] (4) REFUGE: Provides 1,200 fundus images with ground truth segmentations and glaucoma labels, representing the largest dataset for evaluating automated glaucoma assessment methods.[25] (5) ORIGA: Contains 650 retinal images annotated by professionals, focusing on disc and cup segmentation and the cup-to-disc ratio (CDR).[26] (6) RIM-ONE: An open retinal image database for optic nerve evaluation, including 313 fundus images from normal subjects and 172 fundus images from glaucoma patients, with manual segmentation by a glaucoma expert.[27] (7) DRISHTI: consists of 101 fundus images from both normal and glaucomatous eyes, with optic disc and cup segmented by four experts.[28] (8) GAMMA: a multimodal image dataset for glaucoma grading, consisting of fundus and OCT images from 300 patients, annotated with glaucoma grade, macular fovea coordinates, and optic disc/cup segmentation masks.[29]

**Datasets for downstream tasks**

To evaluate the effectiveness of EyeDiff-generated images in optimizing retinal disease diagnosis, we used RETFound as the baseline model and employed nine open-access ophthalmic image datasets.[20] (1) DR diagnosis: The APTOS-2019 (India), IDRiD (India), and MESSIDOR-2 (France) datasets were used, with labels based on the International Clinical Diabetic Retinopathy Severity Scale, indicating five stages: no DR, mild non-proliferative DR (NPDR), moderate NPDR, severe NPDR, and proliferative DR. (2) Glaucoma diagnosis: The PAPILA (Spain) and Glaucoma Fundus (South Korea) datasets were used, including three labels: non-glaucoma, early glaucoma (suspected glaucoma), and advanced glaucoma. (3) Multi-category eye diseases diagnosis: The JSIEC (China), Retina, OCTID and OCTDL datasets were applied. JSIEC consists of 1,000 images, including 39 common fundus diseases and conditions. The Retina dataset has labels for normal, glaucoma, cataract, and retinal diseases. OCTID includes 572 OCT scans labeled as normal, macular hole, age-related macular degeneration (AMD), central serous chorioretinopathy (CSCR), and DR. OCTDL includes labels for normal, AMD, DME, ERM, RVO, etc. (4) Rare diseases diagnosis: Images collected from the Retinal Image Bank between 2019 and 2023 were applied. We created a custom dictionary to select rare disease cases, such as birdshot retinochoroidopathy, cone dystrophy, Stargardt disease, etc. Rare diseases categories can be retrieved from authoritative databases such as the American Academy of Ophthalmology, Orphanet, and the National Organization for Rare Disorders. A total of 2339 images representing 17 rare diseases were included. Supplementary Table 2 demonstrates the details of downstream datasets.

**Well-designed text prompts for guiding image generation**

To obtain semantically rich text guidance, we conducted an information fusion process across multifaceted descriptions of ophthalmic images. We built a custom dictionary to map different expressions of diseases into a structured format using keyword matching with regular expressions. The mapped labels include image modality, lesion or disease annotations, and disease severity (such as mild NPDR, moderate NPDR, etc.). We excluded non-routine retinal examination images, such as histology and pathology images, and conditions with fewer than 50 occurrences. Supplementary Table 3 presents the details of text prompts.

**Model development**

EyeDiff was based on SD v1-5. In our training process, we utilized text prompts as input and the corresponding images as the ground truth to train EyeDiff. Among various text-to-image models, SD has garnered significant attention for its impressive performance in generating high-quality images and its cost-effective fine-tuning. Its denoising process operates in a latent space, akin to diffusion models, resulting in final images that are highly consistent with the input text. This makes SD an excellent tool for text-guided image editing.[21,30] EyeDiff is designed to learn the distribution of ophthalmic images from multiple modalities, conditioned on the corresponding text prompt which contains multifaceted information. During training, images are resized to 512×512 resolution and encoded through an encoder, which turns images into latent representations. The text prompts are encoded through a CLIP text encoder and fed into the UNet backbone of the latent diffusion model via cross-attention. The loss function serves as a reconstruction objective, comparing the noise added to the latent representation with the predictions generated by the UNet. We utilized the AdamW optimizer with a weight decay of 1e-2. The batch size and learning rate were set at 8 and 5e-5, respectively. Each training session was preset to run for a total of 5 epochs. Figure 1(B) presents the algorithm architecture of EyeDiff.

**Image quality evaluations**

**Quantitative evaluation**

We applied novel VQAScore[31] to objectively measure the alignment between generated images and the corresponding texts in downstream tasks. The VQAScore applied a visual-question -answering (VQA) model to generate an alignment score by calculating the probability of a "Yes" answer to the question "Does this figure show '[text]'?". The VQAScore represents a simple and effective metrics that outperforms prior art,[33] demonstrating strong agreement with human judgements. This metric ranges from 0 to 1, with higher scores indicating better alignment.

**Human evaluation**

Fifty images were randomly selected for visual quality evaluation and Turing test by two experienced ophthalmologists (R.C. and X.C.).

**Qualitative Turing test:** The Turing test was conducted using ophthalmic images without the annotations of "real" or "generated". We asked the ophthalmologists (R.C. and X.C.) to determine whether the image was collected from a real patient or was synthesized by our stable diffusion model prompted by a text.

**Visual quality evaluation:** The ophthalmologists evaluated the generated images subjectively using a five-point scale, considering the integrity of generated structures and lesions according to text prompts. The scale is as follows: 1 = The modality and lesion features of the generated image fully match the text prompts; 2 = The modality of the generated image corresponds to the text prompts, and the lesion features mostly align with the text prompts; 3 = The modality of the generated image corresponds to the text prompts, and the lesion features slightly align with the text prompts; 4 = The modality of the generated image corresponds to the text prompts, but the lesion features cannot be generated; 5 = All the text-guided features cannot be generated. To determine the agreement between the ophthalmologists, we calculated Cohen's weighted kappa score. This score ranges from -1 to 1, where values between 0.40 and 0.60 indicate moderate agreement, 0.60 and 0.80 indicate substantial agreement, and 0.80 to 1.00 indicate almost perfect agreement.

**Downstream diagnosis task using generated images**

We evaluated the applicability of EyeDiff-generated images in augmenting minority classes and enhancing overall disease diagnosis. Using the Vision Transformer (ViT), we investigated whether adding EyeDiff-generated images could improve the diagnosis accuracy of multiple retinal diseases. RETFound, a foundation model known for its high accuracy in retinal disease diagnosis, served as our baseline [20] We compared three models, all utilizing the same hyperparameters for classifying retinal diseases: (1) Original real images (RETFound), (2) Original real images + oversampling regular images (Oversample), and (3) Original real images + images generated by our text-guided model (EyeDiff). All models were initialized with pretrained weights from RETFound. Specifically, features from different images were extracted by a ViT large model into 1024-dimensional embeddings. These embeddings were processed through an attention-based feature fusion module, which assigns different weights to each embedding using a multi-head attention mechanism. The weighted features were then aggregated to form a single, fused representation, which was passed through a fully connected layer and a softmax layer to obtain the final diagnosis output.

The performance of retinal disease diagnosis was evaluated through various metrics, such as the area under the receiver operating characteristic curve (AUROC) and the area under the precision-recall curve (AUPR).

**Results**

We finally enrolled 42,048 images from eight datasets for model development and 14,530 images from 10 datasets for downstream common and rare diseases diagnosis. Detailed characteristics of datasets is shown in Table 1.

**Quantitative evaluation**

We finally enrolled 77 categories of texts and the corresponding generated images in downstream datasets for objective quality evaluation. The VQAScore results for text-image alignment in downstream tasks are presented in Table 2. The average VQAScore in datasets for OCT-based disease detection, CFP-based multi-category eye disease diagnosis and multimodal imaging-based rare disease diagnosis is 0.822, 0.776 and 0.670 respectively.

**Human evaluations**

The examples of text prompts and the corresponding generated images are shown in Figure 2. Our model achieved high fidelity in generating detailed structures and lesions in corresponding image modalities according to the text prompts.

**Visual quality evaluation:** Fifty generated images were assessed for visual quality by two experienced ophthalmologists (R.C. and X.C.) using a five-point scale (1 = completely matches the text prompts; 5 = does not match the text prompts). The visual scores were 1.940 ± 1.085 as evaluated by the first grader and 2.080 ± 1.055 by the second grader, with a Kappa value of 0.870.

**Turing test:** Among fifty generated images, ophthalmologists mistook 62% to 66% of them for real images. Most of the generated images shared high similarity with real images. The distinguishing features between generated images and real images included unrealistic colors, enhanced edges of retinal structures or lesions, and high levels of noise.

**EyeDiff improves DR and glaucoma disease diagnosis**

Integrating EyeDiff-generated images with original real images significantly improved the diagnosis accuracy for DR and glaucoma. In the DR diagnosis task using the IDRiD dataset, the AUROC improved from 0.826 (95% CI: 0.821-0.832) at baseline to 0.837 (95% CI: 0.833-0.840) with generated images. The AUPR rose from 0.502 (95% CI: 0.483-0.520) to 0.518 (95% CI: 0.509-0.527), outperforming oversampling-augmented images. For glaucoma diagnosis using the Glaucoma Fundus dataset, the AUROC was 0.950 (95% CI: 0.937-0.964) for original real images, 0.959 (95% CI: 0.945-0.973) for oversampling-augmented images, and 0.959 (95% CI: 0.945-0.973) for EyeDiff-generated images. The AUPR improved from 0.876 (95% CI: 0.841-0.911) to 0.893 (95% CI: 0.855-0.931) with EyeDiff-generated images. These improvements were statistically significant.

**EyeDiff improves multi-class disease diagnosis**

In multi-class disease detection tasks using JSIEC and Retina datasets, the addition of EyeDiff-generated images significantly improved accuracy. For the JSIEC dataset, the AUROC increased from 0.990 (95%CI: 0.989-0.992) to 0.996 (95%CI: 0.995-0.997), and the AUPR rose from 0.887 (95%CI: 0.871-0.891) to 0.967 (95%CI: 0.957-0.978). For the Retina dataset, the AUROC improved from 0.857 (95%CI: 0.831-0.873) to 0.892 (95%CI: 0.867-0.918), and the AUPR increased from 0.720 (95%CI: 0.688-0.761) to 0.779 (95%CI: 0.731-0.826). These differences were statistically significant. In OCT-based disease detection tasks using the OCTID and OCTDL datasets, the addition of generated images also significantly improved accuracy. For the OCTID dataset, the AUROC increased from 0.993 (95% CI: 0.987-0.999) to 0.995 (95% CI: 0.992-0.997), and the AUPR improved from 0.980 (95% CI: 0.967-0.993) to 0.982 (95% CI: 0.969-0.994). For the OCTDL dataset, the AUROC was 0.982 (95% CI: 0.972-0.992) to 0.996 (95% CI: 0.995-0.997) and the AUPR increased

from 0.903 (95% CI: 0.862-0.925) to 0.967 (95% CI: 0.957-0.978). These differences were statistically significant. Table 3 compares the performance of baseline RETFound, EyeDiff, and traditional oversampling on downstream diagnosis tasks.

**EyeDiff improves rare disease diagnosis**

Adding EyeDiff-generated images to the Retina Image Bank significantly enhanced the classification performance of 17 rare diseases. The AUROC was 0.871 (95%CI: 0.863-0.891), 0.893 (95%CI: 0.872-0.923) and 0.919 (95%CI: 0.882-0.931) and the AUPR improved from 0.439 (95%CI: 0.401-0.462) to 0.530 (95%CI: 0.497-0.550).

**EyeDiff-generated images enhance disease classification in minority classes**

Data imbalanced issue exists in these ten downstream datasets and the specific minority classes were demonstrated in Table 4. Supplementing EyeDiff-generated images significantly enhance disease diagnosis performance in these imbalanced classes. (1) IDRiD dataset: The AUROC for mild retinopathy increased from 0.772 (95%CI: 0.733~0.811) to 0.817 (95%CI: 0.780~0.864). (2) APTOS 2019 dataset: The AUROC of severe retinopathy was increased from 0.867 (95%CI: 0.829~0.906) at baseline to 0.914 (95%CI: 0.903~0.934). (3) MESSIDOR2 dataset: The AUROC for proliferative retinopathy increased from 0.960 (95%CI: 0.937~0.988) to 0.980 (95%CI: 0.967~0.994). (4) Glaucoma Fundus dataset: The AUROC of early glaucoma increased from 0.860 (95%CI: 0.827~0.873) to 0.927 (95%CI: 0.919~0.934). (5) PAPILA dataset: The AUROC of glaucoma increased from 0.754 (95%CI: 0.742~0.778) to 0.795 (95%CI: 0.756~0.813). (6) JSIEC dataset: The AUROC of fundus neoplasm increased from 0.739 (95%CI: 0.737~0.756) to 0.875 (95%CI: 0.855~0.877). (7) Retina dataset: The AUROC of cataract increased from 0.951 (95%CI: 0.912~0.954) to 0.961 (95%CI: 0.923~0.977). (8) OCTID dataset: The AUROC of AMD increased from 0.921 (95%CI: 0.890~0.946) to 0.954 (95%CI: 0.947~0.969). (9) OCTDL dataset: The AUROC of RVO increased from 0.992 (95%CI: 0.967~1.011) to 0.993 (95%CI: 0.983~1.051) (10) Retina Image Bank dataset (rare diseases only): The AUROC of optic nerve hypoplasia increased

from 0.701 (95%CI: 0.663~0.713) to 0.774 (95%CI: 0.751~0.792).

**Discussion**

In this study, we developed EyeDiff, a text-to-image diffusion model, to synthesize multimodal ophthalmic images from text prompts. EyeDiff demonstrated robust performance in generating key lesions of various abnormalities based on text prompts, significantly improving the accuracy of classifying both common and rare diseases on top of a well-established foundation model. This data augmentation method outperformed the conventional oversampling methods, offering a promising solution for overcoming challenges in collecting rare, annotated images and facilitating the development of expert-level disease detection models through balanced data generated from simple text cues.

Despite the rarity of individual rare diseases, their collective burden is substantial, affecting over 900 eye abnormalities and leading to lifelong vision impairment in many individuals.[34] Developing accurate DL models for rare eye diseases screening is promising but hindered by the lack of extensive annotated data.. Even RETFound, a powerful generalist model, excels primarily in common disease diagnosis.[20] Previous foundation models like EyeFound and EyeCLIP showed potential in detecting rare eye diseases by learning from unlabeled multimodal retinal images.[35,36] EyeDiff further removes the barriers to collecting rare diseases images, such as macular dystrophy, acute posterior multifocal placoid pigment epitheliopathy, etc. by generating diverse, high-quality images from large-scale multimodal datasets, covering multiple imaging modalities and ocular diseases.

Diffusion models, such as SD, excel in generating realistic images by modeling complex data distributions through forward and reverse diffusion processes.[37] SD integrates text embeddings with image features using a cross-attention mechanism, ensuring generated images accurately reflect text prompts.[21] This approach offers more diverse image resources for DL training compared to image-to-image generation, which is limited by input image quality[22] EyeDiff effectively guides image generation for scarce categories, enhancing model performance and addressing class imbalance.

Generative artificial intelligence (GenAI) technology has been introduced for ophthalmic image generation but are limited to generating a single modality.[15,16,38-41] The diffusion models outperformed GANs in generating images with superior quality, stability, diversity, and controlability.[42] In the current study, both objective measurements and human evaluations confirm the high quality of the generated images. The key lesions of diseases in the generated images closely match the text description, showing great promise for multiple data augmentation tasks. Moreover, we validated the applicability and generalizability of these synthetic images by applying them in downstream worldwide datasets. The addition of text-guided synthetic image significantly enhances the performance of RETFound, even providing additional benefit over traditional data augmentation method. Compared to the original data, EyeDiff can generate privacy-preserving multimodal images, reducing privacy breach risks and removing barriers to cross-center data sharing.

This study has some limitations. The model needs more diverse population data to improve representativeness. Some generated images still exhibit noticeable differences from real ones, necessitating more varied real-world datasets and medical-engineering collaboration to enhance authenticity. Additionally, current text prompts are simplified in this study, optimizing algorithms to handle complex prompts and ensure bias-free descriptions is crucial.

In conclusion, EyeDiff, is a novel generative model for synthesizing multimodal ophthalmic images from natural language prompts. Integrating synthetic images with real data significantly improved diagnosis accuracy for common and rare diseases, outperforming RETFound and traditional oversampling methods. EyeDiff offers an efficient and robust alternative for high-quality, balanced data collection, laying the foundation for developing generalizable and practical disease detection models. Future work should focus on large-scale image-text datasets from multiple regions and optimizing algorithms to enhance the representation and controllability of generated images.

**Funding**

The study was supported by the Start-up Fund for RAPs under the Strategic Hiring Scheme (P0048623) from HKSAR, Global STEM Professorship Scheme (P0046113), and Henry G. Leong Endowed Professorship in Elderly Vision Health. The sponsors or funding organizations had no role in the design or conduct of this research.

**Acknowledgements**

We thank the American Society of Retina Specialists for providing the valuable Retina Image Bank and the InnoHK HKSAR Government for providing valuable supports.

**Conflicts of interest**

The authors declare no competing interest.

**Author contributions**

D.S. conceived the study. D.S. built the deep learning model. D.S., R.C, W.Z. conducted the literature search, analyzed the data. R.C. and X.C. completed visual evaluation. W.Z. performed validation of downstream tasks and quantitative evaluation. R.C. wrote the manuscript. R.C, B.L, P.X. organized figures in this study. M.H., provided the data and facilities. All authors critically revised the manuscript.

**Data and code availability**

The datasets can be accessed by referring to the original paper. The deep-learning model was developed using PyTorch (http://pytorch.org). We trained the model on an NVIDIA V100 card. The code for deep learning model development can be accessed at https://github.com/huggingface/diffusers/tree/main/examples/dreambooth

**Ethics**

We utilized public data for our study, which received approval from the Institutional Review Board of the Hong Kong Polytechnic University.

# References

1. Raimundo, R. & Rosário, A. The Impact of Artificial Intelligence on Data System Security: A Literature Review. *Sensors (Basel, Switzerland)* **21**(2021).
2. Lama, H., *et al.* Severe macular complications in glaucoma: high-resolution multimodal imaging characteristics and review of the literature. *BMC ophthalmology* **23**, 318 (2023).
3. Stino, H., *et al.* Association of Diabetic Lesions and Retinal Nonperfusion Using Widefield Multimodal Imaging. *Ophthalmology. Retina* **7**, 1042-1050 (2023).
4. Rahman, N., Georgiou, M., Khan, K.N. & Michaelides, M. Macular dystrophies: clinical and imaging features, molecular genetics and therapeutic options. *The British journal of ophthalmology* **104**, 451-460 (2020).
5. Ting, D.S.W., *et al.* Artificial intelligence and deep learning in ophthalmology. *The British journal of ophthalmology* **103**, 167-175 (2019).
6. Dong, L., *et al.* Artificial Intelligence for Screening of Multiple Retinal and Optic Nerve Diseases. *JAMA network open* **5**, e229960 (2022).
7. Kihara, Y., *et al.* Policy-Driven, Multimodal Deep Learning for Predicting Visual Fields from the Optic Disc and OCT Imaging. *Ophthalmology* **129**, 781-791 (2022).
8. Özdaş, M.B., Uysal, F. & Hardalaç, F. Classification of Retinal Diseases in Optical Coherence Tomography Images Using Artificial Intelligence and Firefly Algorithm. *Diagnostics (Basel, Switzerland)* **13**(2023).
9. Cen, L.P., *et al.* Automatic detection of 39 fundus diseases and conditions in retinal photographs using deep neural networks. *Nature communications* **12**, 4828 (2021).
10. Aung, Y.Y.M., Wong, D.C.S. & Ting, D.S.W. The promise of artificial intelligence: a review of the opportunities and challenges of artificial intelligence in healthcare. *British medical bulletin* **139**, 4-15 (2021).
11. Gichoya, J.W., *et al.* AI recognition of patient race in medical imaging: a modelling study. *The Lancet. Digital health* **4**, e406-e414 (2022).
12. Shorten, C. & Khoshgoftaar, T.M.J.J.o.B.D. A survey on Image Data Augmentation for Deep Learning. **6**, 1-48 (2019).
13. Vaughan, R. Oversampling in Health Surveys: Why, When, and How? *American journal of public health* **107**, 1214-1215 (2017).
14. Khan, A.A., Chaudhari, O. & Chandra, R. A review of ensemble learning and data augmentation models for class imbalanced problems: Combination, implementation and evaluation. *Expert Systems with Applications* **244**, 122778 (2024).
15. Chen, R., *et al.* Translating color fundus photography to indocyanine green angiography using deep-learning for age-related macular degeneration screening. *NPJ Digit Med* **7**, 34 (2024).
16. Shi, D., *et al.* Translation of Color Fundus Photography into Fluorescein Angiography Using Deep Learning for Enhanced Diabetic Retinopathy Screening. *Ophthalmol Sci* **3**, 100401 (2023).
17. Kugelman, J., *et al.* Data augmentation for patch-based OCT chorio-retinal

segmentation using generative adversarial networks. **33**, 7393 - 7408 (2021).

18. Yoo, T.K., Choi, J.Y. & Kim, H.K. Feasibility study to improve deep learning in OCT diagnosis of rare retinal diseases with few-shot classification. *Medical & biological engineering & computing* **59**, 401-415 (2021).
19. Sonmez, S.C., Sevgi, M., Antaki, F., Huemer, J. & Keane, P.A. Generative artificial intelligence in ophthalmology: current innovations, future applications and challenges. *Br J Ophthalmol* **108**, 1335-1340 (2024).
20. Zhou, Y., et al. A foundation model for generalizable disease detection from retinal images. *Nature* **622**, 156-163 (2023).
21. Rombach, R., et al. High-Resolution Image Synthesis with Latent Diffusion Models. 10674-10685 (2021).
22. Tian, Y., Fan, L., Isola, P., Chang, H. & Krishnan, D.J.A. StableRep: Synthetic Images from Text-to-Image Models Make Strong Visual Representation Learners. **abs/2306.00984**(2023).
23. Gulshan, V., et al. Development and Validation of a Deep Learning Algorithm for Detection of Diabetic Retinopathy in Retinal Fundus Photographs. *Jama* **316**, 2402-2410 (2016).
24. Kulyabin, M., et al. OCTDL: Optical Coherence Tomography Dataset for Image-Based Deep Learning Methods. *Scientific data* **11**, 365 (2024).
25. Orlando, J.I., et al. REFUGE Challenge: A unified framework for evaluating automated methods for glaucoma assessment from fundus photographs. *Medical image analysis* **59**, 101570 (2020).
26. Zhang, Z., et al. ORIGA(-light): an online retinal fundus image database for glaucoma analysis and research. *Annual International Conference of the IEEE Engineering in Medicine and Biology Society. IEEE Engineering in Medicine and Biology Society. Annual International Conference* **2010**, 3065-3068 (2010).
27. Fumero, F., Alayón, S., Sánchez, J.L., Sigut, J.F. & Gonzalez-Hernandez, M.J.t.I.S.o.C.-B.M.S. RIM-ONE: An open retinal image database for optic nerve evaluation. 1-6 (2011).
28. Sivaswamy, J., et al. Drishti-GS: Retinal image dataset for optic nerve head(ONH) segmentation. 53-56 (2014).
29. Wu, J., et al. GAMMA Challenge: Glaucoma grAding from Multi-Modality imAges. **90**, 102938 (2022).
30. Ho, J. Classifier-Free Diffusion Guidance. *ArXiv* **abs/2207.12598**(2022).
31. Lin, Z., et al. Evaluating Text-to-Visual Generation with Image-to-Text Generation. *ArXiv* **abs/2404.01291**(2024).
32. Chen, X., et al. ChatFFA: An ophthalmic chat system for unified vision-language understanding and question answering for fundus fluorescein angiography. *iScience* **27**, 110021 (2024).
33. Hessel, J., Holtzman, A., Forbes, M., Le Bras, R. & Choi, Y. CLIPScore: A Reference-free Evaluation Metric for Image Captioning. *ArXiv* **abs/2104.08718**(2021).
34. Sharma, M. Overcoming challenges in research and development of rare eye diseases. *Indian J Ophthalmol* **70**, 2214-2215 (2022).

35. Shi, D., *et al.* EyeFound: A Multimodal Generalist Foundation Model for Ophthalmic Imaging. *ArXiv* **abs/2405.11338**(2024).
36. Shi, D., *et al.* EyeCLIP: A visual-language foundation model for multi-modal ophthalmic image analysis.   (2024).
37. Kazerouni, A., *et al.* Diffusion models in medical imaging: A comprehensive survey. *Medical image analysis* **88**, 102846 (2023).
38. He, S., *et al.* Bridging the Camera Domain Gap With Image-to-Image Translation Improves Glaucoma Diagnosis. *Transl Vis Sci Technol* **12**, 20-20 (2023).
39. Song, F., Zhang, W., Zheng, Y., Shi, D. & He, M. A deep learning model for generating fundus autofluorescence images from color fundus photography. *Adv Ophthalmol Pract Res* **3**, 192-198 (2023).
40. Shi, D., He, S., Yang, J., Zheng, Y. & He, M. One-shot Retinal Artery and Vein Segmentation via Cross-modality Pretraining. *Ophthalmol Sci* **4**, 100363 (2024).
41. Zhang, W., *et al.* Fundus2Video: Cross-Modal Angiography Video Generation from Static Fundus Photography with Clinical Knowledge Guidance. in *Medical Image Computing and Computer Assisted Intervention – MICCAI* 689-699 (Springer Nature Switzerland, Morocco, 2024).
42. Dhariwal, P. & Nichol, A.J.A. Diffusion Models Beat GANs on Image Synthesis. **abs/2105.05233**(2021).

# Figure legends

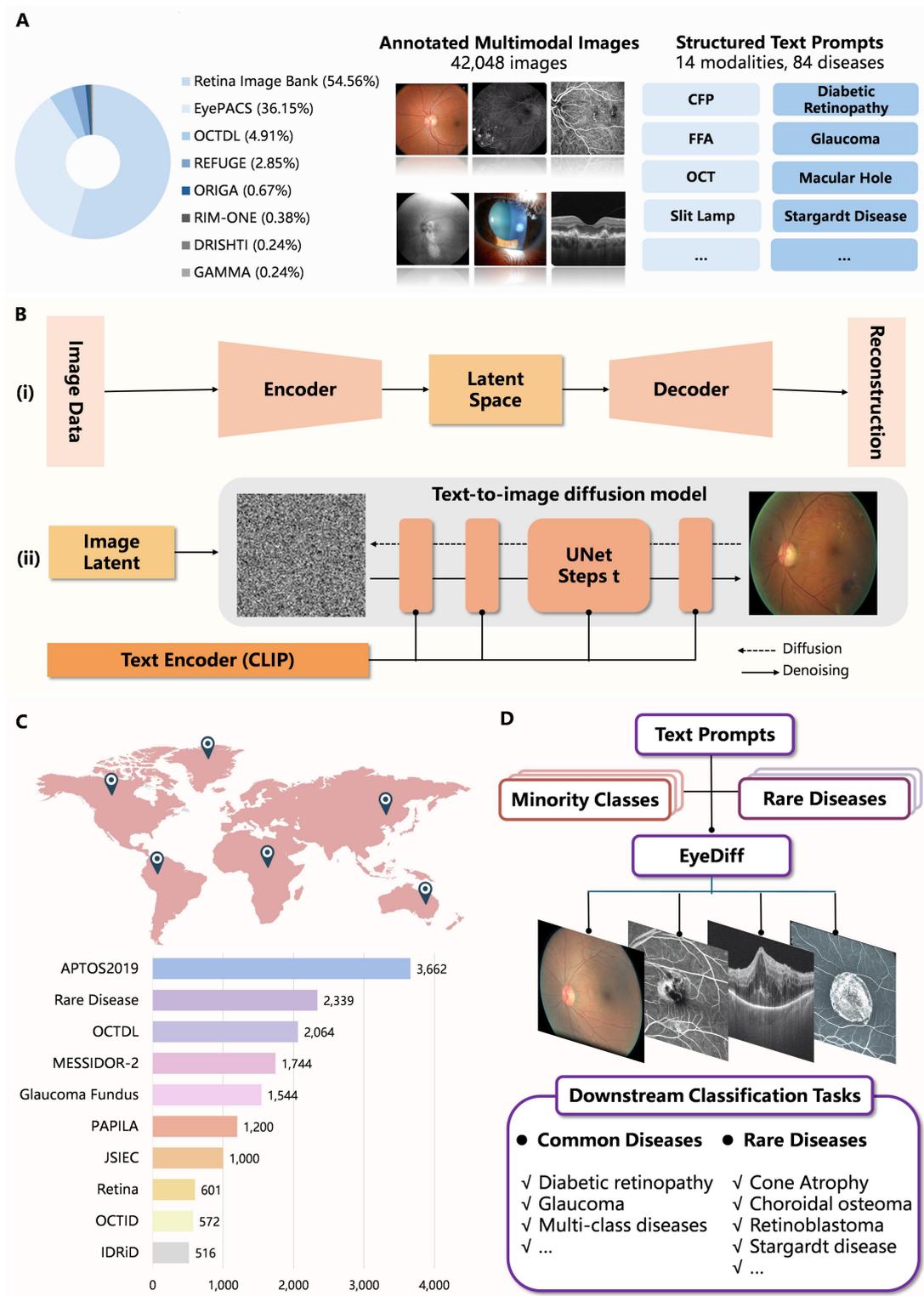

**Figure 1. Flow diagram of this study.** A. Characteristics of training datasets. B. The architecture of the Text-to-image diffusion model. (i) Variational Autoencoder (VAE) architecture for generating ophthalmic image embeddings. Given a retinal image, the encoder transforms it into a latent representation, while the decoder reconstructs the image from this latent. (ii) Text-to-image diffusion model architecture. It is formed by a latent diffusion model (LDM) that consists of time-conditional UNets. The LDM is conditioned on text embeddings derived from the CLIP model. During the training phase, noise is progressively applied to the image latent according to a specified noise scheduler at each timestep t. The LDM learns to reduce this noise given the noisy image latent, the timestep t and the text embedding as inputs. In the inference phase, it begins with completely random noise, guided by a sampling timestep value and a text embedding. Finally, the decoder reconstructs a new synthetic image from the denoised image latent. C. Downstream dataset characteristics. D. Outline of downstream disease classification. CFP=color fundus photography; OCT=optical coherence tomography; FFA=fundus fluorescein angiography

| Text Prompt | Generated Image | Real Reference |
|---|---|---|
| color fundus, glaucoma | 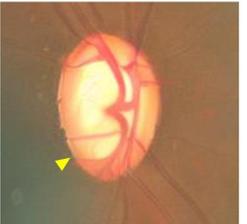 | 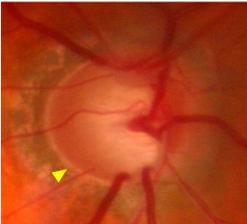 |
| color fundus, retinitis pigmentosa, retinal dystrophy, bone spicules pigmentation pigmentary change | 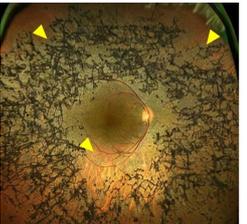 | 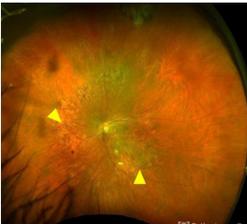 |
| color fundus, severe diabetic retinopathy | 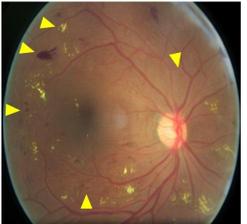 | 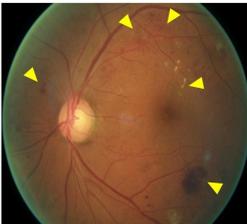 |
| ffa, macular dystrophy, degeneration | 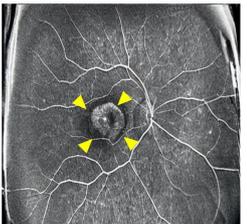 | 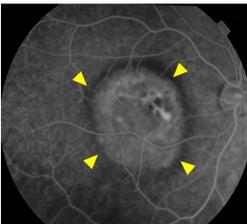 |
| acute posterior multifocal placoid pigment epitheliopathy, white dot syndrome, pigmentary change | 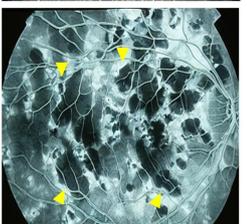 | 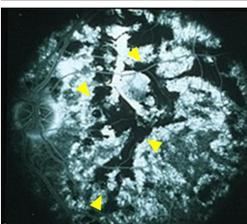 |
| presumed ocular histoplasmosis syndrome, peripapillary atrophy, pigmentary change, scar | 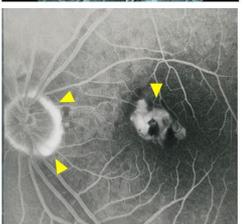 | 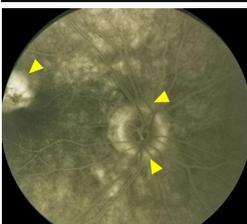 |
| oct, vitreomacular traction, vitreomacular interface disease, macular hole | 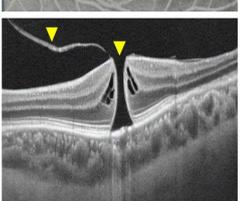 | 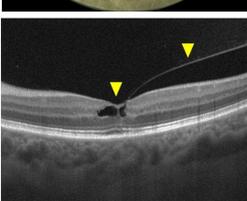 |
| oct, macular edema, retinal vein occlusion | 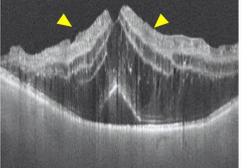 | 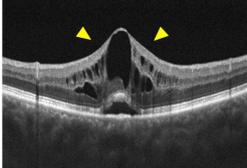 |

**Figure 2. Examples of lesion-preserved generated images using text prompts.** OCT=optical coherence tomography; FFA=fundus fluorescein angiography. The yellow arrow outlines lesions. 1st row: enlarged cup-disc ratio; 2nd row: bone spicules pigmentation and retinal dystrophy; 3rd row: microaneurysm, exudation, retinal hemorrhage, intraretinal microvascular abnormality; 4th row: macular dystrophy and degeneration; 5th row: characteristic lesion of acute posterior multifocal placoid pigment epitheliopathy; 6th row: macular lesion, peripapillary atrophy, pigmentary change; 7th row: vitreomacular traction and macular hole; 8th row: macular edema

**Table 1. Characteristics of datasets**

| Dataset | Image Modality | Disease Category | N (%) |
|---|---|---|---|
| **EyeDiff Development** | | | 42,048 (100.00%) |
| Retina Image Bank | CFP, FFA, ICGA, OCT, FAF, RetCam, B scan, Slit Lamp, SLO, WF-SLO, UWF-SLO, External eye, Red free, OUS | 84 ocular diseases | 22,941 (54.56%) |
| EyePACS | CFP | DR | 15,202 (36.15%) |
| OCTDL | OCT | AMD, DME, RVO, RAO, ERM, VID, Normal | 2,064 (4.91%) |
| REFUGE | CFP | Glaucoma | 1,200 (2.85%) |
| ORIGA | CFP | Glaucoma | 282 (0.67%) |
| RIM-ONE | CFP | Glaucoma, Normal | 158 (0.38%) |
| DRISHTI | CFP | Glaucoma, Normal | 101 (0.24%) |
| GAMMA | CFP, OCT | Glaucoma | 100 (0.24%) |
| **Downstream Validation** | | | 14,530 (100.00%) |
| IDRiD | CFP | DR | 516 (3.55%) |
| APTOS2019 | CFP | DR | 3,662 (25.20%) |
| MESSIDOR-2 | CFP | DR | 1,744 (12.00%) |
| PAPILA | CFP | Glaucoma | 488 (3.36%) |
| Glaucoma Fundus | CFP | Glaucoma | 1,544 (10.63%) |
| JSIEC | CFP | 39 fundus diseases | 1,000 (6.88%) |
| Retina | CFP | Glaucoma, Cataract, Normal | 601 (4.14%) |
| OCTID | OCT | MH, CSCR, AMD DR, Normal | 572 (3.94%) |
| OCTDL | OCT | AMD, DME, RVO, RAO, ERM, VID, Normal | 2,064 (14.20%) |
| Rare Diseases | CFP, FFA, ICGA, OCT, FAF, RetCam, B scan, Slit Lamp, SLO, WF-SLO, UWF-SLO, External eye, Red free, | 17 rare diseases | 2,339 (16.10%) |

OCT=Optical coherence tomography, CFP=Color fundus photography, FFA=Fundus fluorescein angiography, ICGA=Indocyanine green angiography, FAF=Fundus autofluorescence, SLO=Scanning laser ophthalmoscopy, WF-SLO=Widefield SLO UWF-SLO=Ultra-widefield SLO, OUS=Ocular ultrasound, DR=Diabetic retinopathy, AMD=Age-related macular degeneration, DME=Diabetic macular edema, ERM=Epiretinal Membrane, RVO=Retinal vein occlusion, RAO=Retinal artery occlusion, VID=Vitreoretinal interface disease, MH=Macular hole, CSCR=Central serous chorioretinopathy.

**Table 2. The VQAScore results for text-image alignment in downstream tasks.**

| Target dataset | Prompts used in downstream tasks | VQAScore |
|---|---|---|
| **OCT-based disease detection** | optical coherence tomography, retinal artery occlusion | 0.764 |
| | optical coherence tomography, macular hole | 0.789 |
| | optical coherence tomography, vitreomacular interface disease | 0.850 |
| | optical coherence tomography, diabetic macular edema | 0.827 |
| | optical coherence tomography, diabetic retinopathy | 0.762 |
| | optical coherence tomography, epiretinal membrane | 0.888 |
| | optical coherence tomography, central serous chorioretinopathy | 0.841 |
| | optical coherence tomography, normal | 0.794 |
| | optical coherence tomography, retinal vein occlusion | 0.850 |
| | optical coherence tomography, age-related macular degeneration | 0.857 |
| | *Average | 0.822 |
| **CFP-based multi-category eye disease classification** | fundus image, tessellated fundus | 0.861 |
| | fundus image, cataract | 0.824 |
| | fundus image, moderate non-proliferative diabetic retinopathy | 0.841 |
| | fundus image, blur fundus with suspected proliferative diabetic retinopathy | 0.856 |
| | fundus image, retinal disease | 0.842 |
| | fundus image, retinal artery occlusion | 0.636 |
| | fundus image, disc swelling and elevation | 0.795 |
| | fundus image, peripheral retinal degeneration and break | 0.819 |
| | fundus image, blur fundus | 0.805 |
| | fundus image, vessel tortuosity | 0.690 |
| | fundus image, macular hole | 0.830 |
| | fundus image, large optic cup | 0.741 |
| | fundus image, branch retinal vein occlusion | 0.747 |
| | fundus image, non-referable diabetic retinopathy | 0.721 |
| | fundus image, possible glaucoma | 0.882 |
| | fundus image, suspected glaucoma | 0.845 |
| | fundus image, congenital disc abnormality | 0.684 |
| | fundus image, yellow-white spots-flecks | 0.834 |
| | fundus image, severe and proliferative diabetic retinopathy | 0.863 |
| | fundus image, bietti crystalline dystrophy | 0.804 |
| | fundus image, epiretinal membrane | 0.776 |
| | fundus image, vitreous particles | 0.708 |
| | fundus image, central serous chorioretinopathy | 0.844 |
| | fundus image, laser spots | 0.791 |
| | fundus image, glaucoma | 0.827 |
| | fundus image, vogt-koyanagi-harada disease | 0.829 |
| | fundus image, myelinated nerve fiber | 0.717 |
| | fundus image, central retinal vein occlusion | 0.662 |
| | fundus image, optic atrophy | 0.704 |
| | fundus image, silicon oil in eye | 0.816 |
| | fundus image, advanced glaucoma | 0.837 |
| | fundus image, rhegmatogenous retinal detachment | 0.628 |
| | fundus image, maculopathy | 0.767 |
| | fundus image, coloboma | 0.796 |
| | fundus image, massive hard exudates | 0.772 |
| | fundus image, severe hypertensive retinopathy | 0.826 |
| | fundus image, pathological myopia | 0.874 |
| | fundus image, fundus neoplasm | 0.660 |
| | fundus image, cotton-wool spots | 0.829 |
| | fundus image, preretinal hemorrhage | 0.720 |
| | fundus image, early glaucoma | 0.801 |
| | fundus image, dragged disc | 0.805 |
| | fundus image, retinitis pigmentosa | 0.768 |
| | fundus image, fibrosis | 0.795 |
| | fundus image, mild diabetic retinopathy | 0.713 |
| | fundus image, severe diabetic retinopathy | 0.889 |
| | fundus image, normal | 0.274 |
| | fundus image, moderate diabetic retinopathy | 0.819 |
| | fundus image, proliferative diabetic retinopathy | 0.864 |
| | *Average | 0.776 |
| **Multimodal imaging-based rare disease classification** | birdshot retinochoroidopathy | 0.675 |
| | pseudoxanthoma elasticum | 0.629 |
| | familial exudative vitreoretinopathy | 0.683 |
| | macular telangiectasia | 0.751 |
| | central areolar choroidal dystrophy | 0.788 |
| | optic disc pit | 0.761 |
| | retinoblastoma | 0.589 |
| | Stargardt disease | 0.724 |
| | Stargardt disease, cone dystrophy | 0.694 |
| | acute posterior multifocal placoid pigment epitheliopathy | 0.793 |
| | serpiginous choroiditis | 0.648 |
| | serpiginous choroiditis, acute posterior multifocal placoid pigment epitheliopathy | 0.704 |
| | optic nerve hypoplasia | 0.586 |
| | choroidal osteoma | 0.556 |
| | retinopathy of prematurity | 0.493 |
| | congenital hypertrophy of the retinal pigment epithelium | 0.688 |
| | cone dystrophy | 0.652 |
| | choroidal melanoma | 0.565 |
| | retinitis pigmentosa | 0.673 |
| | retinitis pigmentosa, cone dystrophy | 0.757 |
| | *Average | 0.670 |

OCT=Optical coherence tomography, CFP=Color fundus photography, *Average indicates the average VQAScore of the corresponding datasets.

**Table 3. Overall performance comparison of baseline RETFound, EyeDiff, and traditional oversampling on downstream classification tasks**

| Dataset | Model | AUROC (95% CI) | AUPR (95% CI) | P value |
|---|---|---|---|---|
| **IDRiD** | RetFound | 0.826 (0.821, 0.832) | 0.502 (0.483, 0.520) | |
| | EyeDiff | 0.837 (0.833, 0.840) ↑ | 0.518 (0.509, 0.527) ↑ | **0.008*** |
| | Oversample | 0.833 (0.826, 0.841) | 0.516 (0.499, 0.532) | **0.012*** |
| **APTOS 2019** | RetFound | 0.946 (0.941, 0.950) | 0.723 (0.703, 0.745) | |
| | EyeDiff | 0.945 (0.941, 0.948) | 0.714 (0.697, 0.732) | 0.211 |
| | Oversample | 0.944 (0.941, 0.947) | 0.709 (0.689, 0.729) | 0.197 |
| **MESSIDOR2** | RetFound | 0.884 (0.880, 0.887) | 0.669 (0.656, 0.683) | |
| | EyeDiff | 0.879 (0.869, 0.890) | 0.662 (0.631, 0.693) | 0.213 |
| | Oversample | 0.856 (0.840, 0.872) | 0.625 (0.609, 0.641) | 0.187 |
| **Glaucoma Fundus** | RetFound | 0.950 (0.937, 0.964) | 0.876 (0.841, 0.911) | |
| | EyeDiff | 0.959 (0.945, 0.973) ↑ | 0.893 (0.855, 0.931) ↑ | **0.008*** |
| | Oversample | 0.954 (0.940, 0.968) | 0.879 (0.843, 0.915) | **0.009*** |
| **PAPILA** | RetFound | 0.820 (0.788, 0.854) | 0.678 (0.646, 0.709) | |
| | EyeDiff | 0.814 (0.779, 0.848) | 0.676 (0.638, 0.713) | **<0.001*** |
| | Oversample | 0.833 (0.806, 0.859) | 0.724 (0.691, 0.757) | **<0.001*** |
| **JSIEC** | RetFound | 0.990 (0.989, 0.992) | 0.887 (0.871, 0.891) | |
| | EyeDiff | 0.996 (0.995, 0.997) ↑ | 0.967 (0.957, 0.978) ↑ | 0.082 |
| | Oversample | 0.993 (0.990, 0.996) | 0.937 (0.902, 0.951) | 0.133 |
| **Retina** | RetFound | 0.857 (0.831, 0.873) | 0.720 (0.688, 0.761) | |
| | EyeDiff | 0.892 (0.867, 0.918) ↑ | 0.779 (0.731, 0.826) ↑ | **<0.001*** |
| | Oversample | 0.864 (0.852, 0.876) | 0.731 (0.712, 0.750) | **<0.001*** |
| **OCTID** | RetFound | 0.993 (0.987, 0.999) | 0.980 (0.967, 0.993) | |
| | EyeDiff | 0.995 (0.992, 0.997) ↑ | 0.982 (0.969, 0.994) ↑ | **0.017*** |
| | Oversample | 0.993 (0.984, 1.000) | 0.980 (0.958, 0.999) | **0.045*** |
| **OCTDL** | RetFound | 0.982 (0.972, 0.992) | 0.903 (0.862, 0.925) | |
| | EyeDiff | 0.996 (0.995, 0.997) ↑ | 0.967 (0.957, 0.978) ↑ | **<0.001*** |
| | Oversample | 0.994 (0.992, 0.996) | 0.947 (0.928, 0.966) | **<0.001*** |
| **ImageBank** | RetFound | 0.871 (0.863, 0.891) | 0.439 (0.401, 0.462) | |
| | EyeDiff | 0.919 (0.882, 0.931) ↑ | 0.530 (0.497, 0.550) ↑ | **<0.001*** |
| | Oversample | 0.893 (0.872, 0.923) | 0.471 (0.454, 0.501) | **<0.001*** |

CI=confidence interval, AUROC= Area under the receiver operating characteristic curve, AUPR= Area under the precision-recall curve. P values were compared to baseline RETFound. *P<0.05

**Table 4. Comparing the performance of the baseline RETFound model and the use of EyeDiff and oversampling for data augmentation of the minority class in downstream classification tasks**

| Dataset | Minority Class | Model | AUROC (95% CI) | AUPR (95% CI) | P value |
|---|---|---|---|---|---|
| **IDRiD** (n=103) | Mild Retinopathy (n=5) | RetFound | 0.772 (0.733, 0.811) | 0.137 (0.134, 0.144) | |
| | | EyeDiff | 0.817 (0.780, 0.846) ↑ | 0.168 (0.149, 0.178) ↑ | **<0.001*** |
| | | Oversample | 0.804 (0.768, 0.841) | 0.168 (0.138, 0.170) | **<0.001*** |
| **APTOS 2019** (n=1100) | Severe Retinopathy (n=58) | RetFound | 0.867 (0.829, 0.906) | 0.348 (0.323, 0.385) | |
| | | EyeDiff | 0.914 (0.903, 0.934) ↑ | 0.496 (0.482, 0.504) ↑ | **<0.001*** |
| | | Oversample | 0.913 (0.884, 0.920) | 0.372 (0.365, 0.409) | **<0.001*** |
| **MESSIDOR2** (n=526) | Proliferative retinopathy (n=11) | RetFound | 0.960 (0.937, 0.988) | 0.721 (0.712, 0.744) | |
| | | EyeDiff | 0.980 (0.967, 0.994) ↑ | 0.740 (0.732, 0.769) ↑ | **<0.001*** |
| | | Oversample | 0.981 (0.952, 1.007) | 0.724 (0.720, 0.741) | **<0.001*** |
| **Glaucoma Fundus** (n=465) | Early Glaucoma (n=87) | RetFound | 0.860 (0.827, 0.873) | 0.570 (0.561, 0.596) | |
| | | EyeDiff | 0.927 (0.919, 0.934) ↑ | 0.809 (0.791, 0.846) ↑ | **<0.001*** |
| | | Oversample | 0.895 (0.866, 0.915) | 0.791 (0.789, 0.807) | **<0.001*** |
| **PAPILA** (n=98) | Glaucoma (n=14) | RetFound | 0.754 (0.742, 0.778) | 0.391 (0.358, 0.424) | |
| | | EyeDiff | 0.795 (0.756, 0.813) ↑ | 0.401 (0.369, 0.407) ↑ | **<0.001*** |
| | | Oversample | 0.772 (0.739, 0.773) | 0.387 (0.347, 0.407) | **0.017*** |
| **JSIEC** (n=316) | Bietti Crystalline Dystrophy (n=3) | RetFound | 0.990 (0.953, 1.002) | 0.897 (0.881, 0.935) | |
| | | EyeDiff | 0.993 (0.976, 0.994) | 0.931 (0.915, 0.943) | 0.231 |
| | | Oversample | 0.993 (0.97, 1.026) | 0.92 (0.908, 0.939) | 0.521 |
| | Fundus Neoplasm (n=3) | RetFound | 0.739 (0.737, 0.756) | 0.251 (0.243, 0.262) | |
| | | EyeDiff | 0.875 (0.855, 0.877) ↑ | 0.298 (0.298, 0.321) ↑ | **<0.001*** |
| | | Oversample | 0.832 (0.831, 0.853) | 0.255 (0.219, 0.277) | **<0.001*** |
| **Retina** (n=181) | Cataract (n=30) | RetFound | 0.951 (0.912, 0.954) | 0.858 (0.841, 0.894) | |
| | | EyeDiff | 0.961 (0.923, 0.977) ↑ | 0.871 (0.834, 0.889) ↑ | **0.035*** |
| | | Oversample | 0.957 (0.918, 0.974) | 0.873 (0.840, 0.894) | 0.128 |
| **OCTID** (n=175) | Age-related Macular Degeneration (n=32) | RetFound | 0.921 (0.890, 0.946) | 0.937 (0.900, 0.955) | |
| | | EyeDiff | 0.954 (0.947, 0.969) ↑ | 0.946 (0.930, 0.981) ↑ | **0.031*** |
| | | Oversample | 0.940 (0.920, 0.969) | 0.941 (0.940, 0.962) | **0.015*** |
| **OCTDL** (n=621) | Retinal Artery Occlusion (n=7) | RetFound | 0.992 (0.967, 1.011) | 0.979 (0.942, 1.015) | |
| | | EyeDiff | 0.993 (0.983, 1.015) ↑ | 0.981 (0.960, 0.990) ↑ | 0.135 |
| | | Oversample | 0.993 (0.958, 1.012) | 0.982 (0.971, 1.007) | 0.152 |
| **Rare Disease** (n=485) | Optic Nerve Hypoplasia (n=6) | RetFound | 0.701 (0.663, 0.713) | 0.053 (0.045, 0.054) | |
| | | EyeDiff | 0.774 (0.751, 0.792) ↑ | 0.081 (0.049, 0.106) ↑ | **<0.001*** |
| | | Oversample | 0.731 (0.693, 0.736) | 0.060 (0.042, 0.093) | **<0.001*** |

CI=confidence interval, AUROC=Area under the receiver operating characteristic curve, AUPR=Area under the precision-recall curve, n= Number of Images. P values were compared to baseline RETFound. *P<0.05. The results displayed in the table are validated on the test set.